\documentclass[12pt]{article}

\begin{document}

\title{The Theory of Stochastic Space-Time. \\1. Gravitation as a Quantum Diffusion\footnote{Published in: ''Z.Zakir (2003)
 textit\{Structure of Space-Time and Matter.\} CTPA, Tashkent.''}}
\author{Zahid Zakir\thanks{E-Mail: zahid@in.edu.uz}\\Centre for Theoretical Physics and Astrophysics\\
P.O.Box 4412, Tashkent, 700000 Uzbekistan}
\date{December 31, 1998;\\
Revised: October 17, 2003. }

\maketitle
\begin{abstract}
The Nelson stochastic mechanics of \textit{inhomogeneous} quantum diffusion in
flat spacetime with a tensor of diffusion can be described as a
\textit{homogeneous} one in a Riemannian manifold where this tensor of
diffusion plays the role of a metric tensor. It is shown that the such
diffusion accelerates both a sample particle and a local frame such that their
mean accelerations do not depend on their masses. This fact, explaining the
principle of equivalence, allows one to represent the curvature and
gravitation as consequences of the quantum fluctuations. In this diffusional
treatment of gravitation it can be naturally explained the fact that the
energy density of the instantaneous Newtonian interaction is negative defined.
\end{abstract}

\section{Introduction}

Two basic phenomena of physics - the gravitation and quantum fluctuations -
both depend only on masses of objects, and both have a geometric nature. The
geometrical origin of gravitation is well known, while up to now Nelson's
discovery of the stochastic geometry of spacetime \cite{Ne} does not accepted
as one of fundamental concepts of physics. The stochastic mechanics is unique
version of quantum mechanics where the quantum fluctuations are represented as
manifestations of stochastic geometry of spacetime with a \textit{constant}
diffusion coefficient $\nu_{0}=\hbar/2m$ (see also the review \cite{Bla}).

It is natural to consider a more general case of spacetimes with
\textit{inhomogeneous} diffusion by a tensor of diffusion $\nu^{ab}(x,t)$. In
the paper it will be shown that the stochastic structure of the spacetime with
the such tensor of diffusion induces a nontrivial metric $g^{ab}(x,t)$ and a
curvature. This means that gravitation can be treated as a quantum diffusional effect.

The fact that the quantum fluctuations and gravitation are not independent
phenomena, leads to the interesting solution of the problem of the
gravitational energy. It is well known that the energy density of the
instantaneous Newtonian interaction, as only attractive one, is negative
defined. In the diffusional treatment of gravity this property can be
naturally explained.

\section{Stochastic mechanics of the homogeneous diffusion}

Let we have Nelson's diffusion \cite{Ne} of a non-relativistic particle of
mass $m$ in euclidean space $R^{n}$:
\begin{equation}
dx_{\pm}^{a}(t)=b_{\pm}^{a}(x,t)dt+dw_{\pm}^{a}(t),
\end{equation}
with:
\begin{equation}
\lim\limits_{\Delta t\rightarrow0}E[\Delta w_{\pm}^{a}(t)\mid x^{a}(t)]=0,
\end{equation}%
\begin{equation}
\lim\limits_{\Delta t\rightarrow0}E[\Delta w_{\pm}^{a}(t)\Delta w_{\pm}%
^{b}(t)\mid x^{a}(t)]=\pm2\nu_{0}\delta^{ab}dt.
\end{equation}

Here the diffusion coefficient is taken as $\nu_{0}=\hbar/2m$, where $\hbar$
is the Planck constant, $\Delta w_{\pm}^{a}(t)=w^{a}(t\pm\Delta t)-w^{a}(t)$,
and the limit $\Delta t\rightarrow0$ should be taken only after the
calculation of the conditional expectations $E[...\mid x^{a}(t)]$. The drifts:%

\begin{equation}
b_{\pm}^{a}(x,t)=\lim\limits_{\Delta t\rightarrow0}E\left[  \frac{\Delta
x_{\pm}^{a}(t)}{\pm\Delta t}\mid x^{a}(t)\right]  ,
\end{equation}
allow one to define the current and osmotic velocities:
\begin{eqnarray}
v^{a}  &  =&\frac{1}{2}(b_{+}^{a}+b_{-}^{a}),\\
u^{a}  &  =&\frac{1}{2}(b_{+}^{a}-b_{-}^{a}).
\end{eqnarray}

The acceleration is defined as:
\begin{eqnarray}
a^{a}(x,t)  &  =&\frac{1}{2}(D_{+}D_{-}+D_{-}D_{+})x^{a}(t)\nonumber\\
&  =&\frac{\partial v^{a}}{\partial t}+(\mathbf{v\nabla})v^{a}-(\mathbf{u\nabla
})u^{a}-\nu\Delta u^{a},
\end{eqnarray}
where the stochastic derivatives $D_{\pm}$ are:
\begin{eqnarray}
D_{\pm}f  &  =&\lim\limits_{\Delta t\rightarrow0}E\left[  \frac{f[x(t\pm\Delta
t),t\pm\Delta t]-f(x,t)}{\pm\Delta t}\mid x^{a}(t)\right]\nonumber\\
&  =&\frac{\partial f}{\partial t}+(\mathbf{b}_{\pm}\mathbf{\nabla})f\pm
\nu\Delta f.
\end{eqnarray}

The mean trajectories of free classical particles by the current velocity
$\mathbf{v}$ are diffusional geodesic lines on $R^{n}$. The equation of motion
for them in the external field is the Newton equation:
\begin{equation}
m\mathbf{a}=-\mathbf{\nabla} V.
\end{equation}

The mean value of the acceleration is:
\begin{equation}
E[a^{b}(t)]=\int dxa^{b}(x,t)\rho(x,t)=\int dx\rho\left[  \frac{\partial
v^{b}}{\partial t}+(\mathbf{v\nabla})v^{b}\right]  \rho,
\end{equation}
which for the Newtonian potential $\varphi_{N}$ gives:
\begin{equation}
E\left[  \frac{\partial v^{b}}{\partial t}+(\mathbf{v\nabla})v^{b}\right]
=-\nabla^{b}\varphi_{N}.
\end{equation}
Therefore, the mean acceleration of the sample particle in the static
gravitational field $\varphi_{N}$ does not depend on the mass of this particle
(the principle of equivalence).

\section{Stochastic mechanics of inhomogeneous diffusion.}

Let us consider in $R^{n}$ a general diffusion with the tensor of diffusion:
\begin{equation}
\nu_{ab}(x,t)=\nu_{0}\gamma_{ab}(x,t),
\end{equation}
where $\gamma_{ab}$ is the normalized tensor of diffusion, which in the case
of Nelson's homogeneous diffusion has been taken as: $\gamma_{ab}=\delta_{ab}%
$. The mean trajectories of particles in this general case are not geodesic
lines on $R^{n}$, but contain some deviations from the geodesics. For the
description of the general diffusion we introduce \textit{the curvilinear
coordinates }$x^{i}(x^{a},t)$\textit{\ and basis vectors }$e_{i}^{a}%
$\textit{\ along the mean trajectories }of the free particles. Then we have
for the local physical coordinates of the sample particle at a point $M$:
\begin{eqnarray}
dx^{i}(M,t)  &  =&e_{a}^{i}(M,t)dx^{a}(M,t),\\
dx_{\pm}^{a}(t)&=&b_{\pm}^{a}(x,t)dt+dw_{\pm}^{a}(t),
\end{eqnarray}
where $e_{i}^{a}e_{i}^{b}=\delta^{ab}$, $e_{i}^{a}e_{j}^{a}=g_{ij}$, and
$g^{ab}=e_{i}^{a}e_{j}^{b}g^{ij}$ is the metric of flat spacetime $R^{n}$,
$g^{ij}(x,t)$ is the metric tensor of the manifold $M$ with the curvilinear
coordinates $x^{i}$, formed by the mean trajectories of the current velocity
$v^{i}(x,t)$ at the free diffusion:%

\begin{equation}
dx_{\pm}^{i}(t)=b_{\pm}^{i}(x,t)dt+d[e_{a}^{i}(x,t)w_{\pm}^{a}(t)].
\end{equation}

Then the conditional expectations can be defined in terms of these curvilinear
coordinates only for small temporal intervals $\Delta t$ and along a
piece-wise smooth curve approximating random curves near the point $M $. The
corresponding tensor of diffusion is defined as:%

\begin{equation}
\nu^{ij}(x,t)=\frac{1}{2}\lim\limits_{\Delta t\rightarrow0}E_{\gamma}\left[
\frac{\Delta x_{\pm}^{i}(t)\Delta x_{\pm}^{j}(t)}{\pm\Delta t}\mid
x^{i}(t)\right]  =\nu_{0}\gamma^{ij}(x,t).
\end{equation}

If we introduce new type of the parallel transport of tensors - \textit{the
stochastic parallel transport }in flat spacetime \textit{along the mean
trajectory of the free drift:}%

\begin{equation}
de_{a}^{i}(x,t)=-\Gamma_{ml}^{i}e_{a}^{l}dx^{m}(t)-\frac{1}{2}d[\Gamma
_{ml}^{i}e_{a}^{l}]dx^{m}(t).
\end{equation}
The such inhomogeneous diffusion can be described as a homogeneous one in the
Riemannian manifold with a constant diffusion coefficient $\nu_{0}=const $ and
the metric tensor $g^{ij}(x,t)$ by the identification $\gamma^{ij}%
(x,t)=g^{ij}(x,t).$ The stochastic mechanics with the tensor of diffusion then
can be treated as quantum mechanics in Riemannian manifold. Therefore, one can
use the well known formulas of the stochastic mechanics in curved manifolds
\cite{Gu1}.

The stochastic derivatives have the form:%

\begin{equation}
(D_{\pm}F)^{i}(x,t)=\frac{\partial F^{i}}{\partial t}+(\mathbf{b}_{\pm
}\mathbf{\nabla})F^{i}\pm\nu_{0}(\Delta_{DR}F)^{i},
\end{equation}
where $\nabla-$ is the Laplace-Beltrami operator in the curved manifold, and%

\begin{equation}
(\Delta_{DR}F)^{i}=\Delta F^{i}+R_{j}^{i}F^{j},
\end{equation}
is the Laplace-de Rham operator, $\Delta=\mathbf{\nabla\nabla},$ and
$R_{j}^{i}$ is the Ricci tensor.

From the expression for the acceleration:%

\begin{equation}
a^{i}(x,t)=\frac{1}{2}(D_{+}b_{-}+D_{-}b_{+})^{i}(x,t)=-\frac{1}{m}\nabla
^{i}V,
\end{equation}
one can obtain the equations:%

\begin{equation}
\frac{\partial v^{i}}{\partial t}+(\mathbf{v\nabla})v^{i}-(\mathbf{u\nabla
})u^{i}-\nu_{0}\Delta u^{i}-\nu_{0}R_{j}^{i}F^{j}=-\frac{1}{m}\nabla^{i}V.
\end{equation}

The continuity equation for the probability density $\rho(x,t)$%
\begin{equation}
\frac{\partial\rho}{\partial t}+\nabla_{i}(\rho v^{i})=0,
\end{equation}
gives $u^{i}=\nu_{0}(\nabla^{i}\rho)/\rho.$ Then by means Nelson's assumptions
$mv^{i}=\nabla^{i}S,$ $\nu_{0}=\hbar/2m,$ and%

\begin{equation}
\psi(x,t)=\sqrt{\rho(x,t)}\exp[iS(x,t)/\hbar],
\end{equation}
where $S(x,t)$ is some function, $\psi(x,t)$ is wave function, we obtain the
Schr\"{o}dinger equation for particle's motion in the stochastic space with
the tensor of diffusion \cite{Gu1}:
\begin{equation}
i\hbar\frac{\partial\psi}{\partial t}=-\frac{\hbar^{2}}{2m}\Delta\psi+V\psi.
\end{equation}

\section{The diffusion induced gravity}

The mean value of acceleration does not contain the terms with the osmotic
velocity $u^{i}$ and although no any external field, nevertheless, there
appears \textit{a proper acceleration} due to the presence of the derivatives
of the metrics in the Laplace-Beltrami operator:
\begin{equation}
E\left[  \frac{\partial v_{i}}{\partial t}+(\mathbf{v\nabla})v_{i}\right]  =0,
\end{equation}
which means that:
\begin{equation}
E\left[  \frac{\partial v_{i}}{\partial t}+v^{j}\partial_{j}v_{i}\right]
=E\left[  \Gamma_{ij}^{k}v^{j}v_{k}\right]  .
\end{equation}

This \textit{diffusion induced mean acceleration does not contain a dependence
on the mass} of the particle, i.e. we have an analog of the equivalence
principle as for gravitation.

The stochastic mechanics has been naturally generalized to the case of
relativistic particle \cite{Gu2}. In this case we obtain non-trivial metrics
for spacetime, which we can interpret as diffusion induced gravity. In other
words, inhomogeneous diffusion leads to the acceleration of the particle
exactly the same as some effective gravitational field.

It is very important that independence of the acceleration on the mass of the
sample particle leads also \textit{to the same proper acceleration of
macroscopic objects - basises of reference frames.} The acceleration of the
reference frame means the appearance of non-trivial metrics and non-zero
curvature. Einstein's equations for the curvature and the metric tensor then
can be represented as equations for the tensor of diffusion.

The identification of the metric structure of spacetime with the general
diffusion leads to the interpretation of gravitation as a secondary effect of
the quantum fluctuations.

In the stochastic interpretation of gravitation the equations for the metrics
$g_{ij}(x,t)$ can be obtained from the diffusion equations. The gravitation is
the energy-momentum effect and we deal with some analog of a
\textit{thermodiffusion}, where the diffusional flow $\mathbf{j}$ is directed
opposite to the gradient of the vacuum energy density in space:
\begin{equation}
\mathbf{j}=-\alpha\mathbf{\nabla}\rho_{(0)},
\end{equation}
where $\alpha$ is some constant. Let a sample particle is at rest at very
large distance from the source where the energy density of the vacuum is
$\rho_{(0)1}$. Due to the inhomogeneous diffusion, the particle will move to
the massive source, where the energy density is $\rho_{(0)2}$. The diffusion
leads to the drift of the sample particle from the intensive fluctuating
region of the background to the lower intensity region. Therefore, for
\textit{the attractive diffusion} there must be negative difference of the
energy densities: $\rho_{(0)2}<\rho_{(0)1}.$

Therefore, if we interpret the gravitation as the such diffusion process, the
energy density corresponding to the gravitational interaction must be negative
defined. The fact that the energy density of the instantaneous Newtonian
interaction is negative defined, is well know, and we can treat this property
of gravitation as one of evidences of its diffusional origin. Here matter's
energy-momentum density tensor $T_{ij}$ plais the role of a source for the
deformations of the stochastic structure of spacetime.

Thus, we have some \textit{duality} between the deformations of the stochastic
structure and the Riemannian structure of spacetime.

\section{Conclusions}

The main result of the paper is the fact, that the quantum fluctuations
induces the gravitation and that the gravitation can be treated as a
macroscopic remainder of inhomogeneous quantum fluctuations. This result is
very important not only for the understanding of a physical nature of
gravitation, but for the strategy of the construction of unified theories
also. If, in the stochastic treatment, the gravitation and quantum
fluctuations are not independent phenomena, then it becomes clear that the
unified theories must lead to an explanation of this fundamental fact.

In the paper \cite{Za} we will show that the stochastic treatment of the
quantum theory and gravitation are not only some hypotheses, but can be
considered as consequences of some general principles of invariance.

\end{document}